\documentstyle[osa,psfig,epsf]{revtex}
\newcommand{\beq}{\begin{equation}}
\newcommand{\eeq}[1]{\label{#1} \end{equation}}
\newcommand{\beqar}{\begin{eqnarray}}
\newcommand{\eeqar}[1]{\label{#1} \end{eqnarray}}
\newcommand{\g}{\gamma}
\newcommand{\insertplot}[1]{
     \centerline{\psfig{figure={#1},height=9.0cm}}
}
\begin{document}
\begin{center}
{\large\bf
Non-ideal Particle Distributions from Kinetic Freeze Out Models 
}
\bigskip

Cs. Anderlik,$^1$
Zs.I. L\'az\'ar,$^1$
V.K. Magas,$^1$\\  
L.P. Csernai,$^{1,2}$ 
H. St\"ocker$^3$
and W. Greiner$^3$
\end{center}
\bigskip

{\normalsize\noindent
\hspace*{-8pt}$^1$ Section for Theoretical Physics, Department of Physics\\
University of Bergen, Allegaten 55, 5007 Bergen, Norway\\[0.2ex]
\hspace*{-8pt}$^2$ KFKI Research Institute for Particle and Nuclear Physics\\
P.O.Box 49,  1525 Budapest, Hungary\\[0.2ex] 
\hspace*{-8pt}$^3$ Institut f\"ur Theoretische Physik, Universit\"at Frankfurt\\
Robert-Mayer-Str. 8-10, D-60054 Frankfurt am Main, Germany
}

\bigskip

{\bf Abstract:} In fluid dynamical models the freeze out of particles across
a three dimensional space-time hypersurface is discussed. The calculation of
final momentum distribution of emitted particles is described for freeze out
surfaces, with both space-like and time-like normals, taking into account
conservation laws across the freeze out discontinuity.

\section{Introduction}

The freeze out of particle distributions is an essential part of continuum or
fluid dynamical reaction models.  From the point of view of observable
consequences this is one of the most essential parts of the model. On the
other hand this step is not based on fluid dynamical principles and governed
by a large variaty of ad hoc assumptions.  The freeze out can be considered
as a discontinuity across a hypersurface in space-time.

The general theory of discontinuities in relativistic flow was not worked out
for a long time, and the 1948 work of A.  Taub \cite{Ta48} discussed
discontinuities across propagating hypersurfaces only (which have a
space-like normal vector, $d\sigma^\mu d\sigma_\mu = - 1$).  Events happening
on a propagating, (2 dimensional) surface belong to this category.

Another type of change in a continuum is an overall sudden change in a finite
volume. This is represented by a hypersurface with a time-like normal,
$d\sigma^\mu d\sigma_\mu = + 1$, called confusingly both space-like and
time-like surface in the literature. In 1987 Taubs approach was generalized
to both types of surfaces,\cite{Cs87} making it possible to take into account
conservation laws exactly across any surface of discontinuity in relativistic
flow.  This approach also eliminates the imaginary particle currents arising
from the equation of the Rayleigh line. When the EoS is different on the two
sides of the freeze out front these conservation laws yield changing
temperature, density, flow velocity across the front.

In fact the freeze out surface is an idealization of a layer of finite
thickness where the frozen out particles are formed, and the interactions in
the matter become gradually negligible.  The dynamics of this layer can be
described in different kinetic models or four-volume emission
models.\cite{AC98} The zero thickness limit of such a layer is the idealized
freeze out surface.
 
The invariant number of conserved particles (world lines)
crossing a surface element, $d\sigma^\mu$, is  
$
dN = N^\mu \ d\sigma_\mu \ ,
$ 
and the total number of all the particles crossing the
FO hyper-surface, $S$, is
$
N = \int_S N^\mu \ d\sigma_\mu \ .
$ 
This total number, $N$, and the total energy and momentum are of course the
same at both sides of the freeze out surface.  
If we insert the kinetic definition of $N^\mu$ 
$$
N^\mu = \int \frac{d^3p}{p^0}\ p^\mu\ f_{FO}(x,p;T,n,u^\nu) \ ,
$$
into the integral over the freeze out hypersurface, $S$, 
we obtain the Cooper-Frye formula:\cite{CF74}
\beq
E\frac{dN}{d^3p} = \int  f_{FO}(x,p;T,n,u^\nu) \ p^\mu d\sigma_\mu\ ,
\eeq{e-cf}
where $f_{FO}(x,p;T,n,u^\nu)$ is the post FO phase space distribution of
frozen-out particles which is not known from the fluid dynamical model.
Problems usually arise from the bad choice of this distribution.  First of
all, to evaluate measurables we have to use the correct parameters of the
matter after the FO discontinuity!

If we know the pre freeze out baryon current and energy-momentum tensor, 
$N_0^\mu$ and $T_0^{\mu\nu}$, 
we can calculate locally, across a surface element of normal vector
$d\sigma^\mu$
the post freeze out quantities, 
$N^\mu$ and $T^{\mu\nu}$, 
from the relations \cite{Ta48,Cs87}:
$
[N^\mu\ d\sigma_\mu] = 0
$ 
and
$
[T^{\mu\nu}\ d\sigma_\mu] = 0 ,
$
where $[A]\equiv A - A_0$.  In numerical calculations the local freeze out
surface can be determined most accurately via self-consistent
iteration.\cite{Bu96,NL97} This fixes the parameters of our post FO momentum
distribution, $f_{FO}(x,p;T,n,u^\nu)$.

For example we can illustrate the effect of conservation laws
for a situation where the frozen out
matter is {\em massless, baryonfree Bose gas}.
Then,
the conservation laws across the freeze-out surface with 
{\em timelike} normal vector 
$d\sigma^\mu$ are
$
[{T^{\mu\nu}d\sigma_\nu}]=0\,,
$
In the most general (three dimensional) case 
there are four parameters to be determined from the conservation laws: The
final, post FO temperature, $T$, and three components of the velocity, $u$.
The energy-momentum tensor on the pre
freeze-out side, and the normal to the surface are given. 
The post freeze-out  energy-momentum tensor is of the form
$
T^{\mu\nu}=$ $(e+p)u^\mu u^\nu-pg^{\mu\nu}
$,
where the energy density, pressure, and temperature are connected by the EoS:
$e=\sigma_{SB} T^4=3p$, where $\sigma_{SB}$ is the Stefan-Boltzmann constant.
Then
$
T^{\mu\nu}=$ $(e+p)u^\mu u^\nu-pg^{\mu\nu}
$,
 can be written as a vector equation:
\begin{equation}
(4u^\mu u^\nu d\sigma_\nu- d\sigma^\mu)=x a^\mu\,,
\label{equations1}
\end{equation}
where 
$$
x=\left(\frac{1}{3}\sigma_{SB} T^4\right)^{-1}\quad,\quad
a^\mu=T_0^{\mu\nu} d\sigma_\nu\,.
$$
Taking the normal projection of (\ref{equations1}) 
and the norm of the four-velocity, $u^\mu$, the solution for the 
four quantities we are looking for will bei given by:
\begin{equation}
x=
\frac{a^\mu d\sigma_\mu+\sqrt{(a^\mu d\sigma_\mu)^2+3a^\mu a_\mu}}
{a^\mu a_\mu}
\quad,\quad 
u^\mu=\frac{x a^\mu+d\sigma^\mu}{2\sqrt{x a^\mu d\sigma_\mu+1}}\,.
\label{results1}
\end{equation}

\paragraph{Idealized freeze out across propagating discontinuities.} 
One can go a step further in the study of freeze out process.  We usually
assume that the pre freeze out momentum distribution as well as the post
freeze out distribution are both local thermal equilibrium distributions
boosted by the local collective flow velocity on the actual side of the
freeze out hypersurface, although, the post freeze out distribution need not
be a thermal distribution!

The case of freeze out across a hypersurface with space-like normal
shows this clearly
because $p^\mu$ is time-like and $d\sigma^\mu$ is space-like, thus 
$p^\mu d \sigma_\mu $
can  be both positive and negative. I.e., $p^\mu$
may point now both in the post- and pre- FO directions.
Thus the integrand in the above integral (\ref{e-cf}) may change sign
in the integration domain,
and this indicates that part of the distribution contributes to a
current going back, into the front while another part is coming out of the
front.  On the pre freeze out side 
$p^\mu$ is unrestricted and 
$p^\mu d \sigma_\mu $
may really have both signs, because we may assume that the freeze out front
has a certain thickness\cite{BB95}, and due to internal rescatterings inside
this front a current is fed back to the pre freeze out side to maintain the
thermal equilibrium there.

On the post freeze out side, however, the distribution,
$
f^*_{FO}(x,p;d\sigma^\mu)  
$
must vanish for momentum four-vectors, $p^\mu$, which point back
into the pre FO direction, i.e. do not satisfy the
condition, $p^\mu d \sigma_\mu > 0$.\cite{Si89,Bu96} 
Thus, this distribution cannot be
a J\"uttner- or other ideal gas distribution.% 
\footnote{
Note, that the contravariant normal when becomes space-like, $d\sigma^\mu$,
should point into the pre-FO direction to satisfy the  
condition, $p^\mu d \sigma_\mu > 0$,
while the covariant normal, $d\sigma_\mu$, always points into the post-FO
direction! Thus, the direction of the contravariant normal
$d\sigma^\mu$, in the Cooper-Frye formula goes continuously over from
pointing to the pre-FO direction to pointing to the post-FO direction
while the covariant normal of the FO surface stays directed
always in the post-FO direction when it goes continuously over from 
time-like to space-like.} 

Nevertheless, the above conservation laws, have to be satisfied, even if the
post freeze out distribution is not a local thermal distribution.  Since, the
kinetic definitions of the energy-momentum tensor and conserved current(s)
are reliably applicable, the conservation laws across a small element of the
freeze out front take the form:
\beq
    \int_S  \left(\int \frac{d^3p}{p^0}\  \ 
    f^*_{FO}(x,p;T,n,u^\nu,d\sigma^\gamma)\  
    p^\mu \right)\ d\sigma_\mu  =
    \int_S  N^{\mu}_0(x) \  d\sigma_\mu  \ ,
\eeq{efo7}
\beq
 \int_S  \left(\int \frac{d^3p}{p^0}\  \
    f^*_{FO}(x,p;T,n,u^\sigma,d\sigma^\gamma)\  
    p^\mu p^\nu \right)\ d\sigma_\mu  =
 \int_S  \ T_0^{\mu \nu}(x)\  d\sigma_\mu  \ .
\eeq{efo8}
Here, the matter is characterized by $T_0^{\mu\nu}$ and $N_0^\mu$ on the
pre freeze out side of the front.

The construction of the post freeze out distribution, $f^*_{FO}$, is a
problem in the case of freeze out fronts with space-like
normal. 

For cut J\"uttner distribution 
the conserved currents 
were evaluated in
ref. \cite{AC98}.
Thus, if we know the 5 parameters of the pre FO flow and 
the local freeze out surface from kinetic considerations, then assuming
that the post FO distribution, $f^*_{FO}(p,x)$, is a cut J\"uttner
distribution, we can completely determine the parameters of the post FO
matter from the conservation laws (\ref{efo7},\ref{efo8}).
Although, this way we would formally satisfy the conservation laws
and we would eliminate the particle current pointing
back to the pre FO matter, the strange shape of the cut J\"uttner 
distribution makes it difficult to accept it as a physical
post FO momentum distribution.

\section{Freeze out distribution from kinetic theory}
\label{six}

Following the ideas introduced in ref. \cite{AC98}
we can calculate the kinetic freeze out distribution based on
four-volume emission models.  The proposed model, on the other
hand, requires extended numerical calculation, so here we
intend to study some overly simplified models, which might
give us some hints about the expected shape of post freeze out distributions.
 
The freeze-out will turn out to be  an exponential
process, and  after about three mean free pathes the amount of
interacting matter reduces to 5 per cent.  Thus, the sharp 
freeze out layer turns out to be an over-idealization  of kinetic freeze out
in heavy ion reactions, while it is applicable on more macroscopic scales
like in astrophysics.%
\footnote{
On the other hand, 
if kinetic freeze out coincides with a rapid phase transition, like in the
case of rapid deconfinement transition of supercooled quark-gluon plasma,
the short freeze out hypersurface idealization may still be applicable
even for heavy ion reactions. It is, however, beyond the scope of this
work to study the freeze out dynamics and kinetics in this latter case.
}

Let us first demonstrate the kinetic model for a drastically oversimplified
situation of a plane FO surface. Let us assume an infinitely long tube
with its left half ($x<0$) filled with nuclear mater and in the right 
vacuum is maintained. We can remove the dividing wall at $t=0$, and then 
the matter will expand into the vacuum. By continuously removing
particles at the right end of the tube and supplying particles on the 
left end, we can establish a stationary flow in the tube, where the
particles will gradually freeze out in an exponential rarefaction wave
propagating to the left.  We can move with this front, so
that we describe it from the reference frame of the front (RFF).

In this frame, we have a stationary supply of equilibrated matter from the
left, and a stationary rarefaction front on the right, $x>0$.
We can describe the freeze out kinetics on the r.h.s. of the tube
assuming that we have two components of our momentum 
distribution,
$f_{free}(x,\vec{p})$  and
$f_{int}(x,\vec{p})$. However, we assume only that at $x=0$ $f_{free}$
vanishes exactly and $f_{int}$ is an ideal J\"uttner distribution
(supplied by the inflow of equilibrated matter), while $f_{int}$ gradually
disappears and $f_{free}$ gradually builds up as $x$ tends to infinity.
We do not assume a priory that $f_{int}(x,\vec{p})$ is an ideal J\"uttner 
distribution for all $x$, 
so we will have different FO results depending
on the assumed FO mechanism.

%   fig. 1

\begin{figure}[htb]

\insertplot{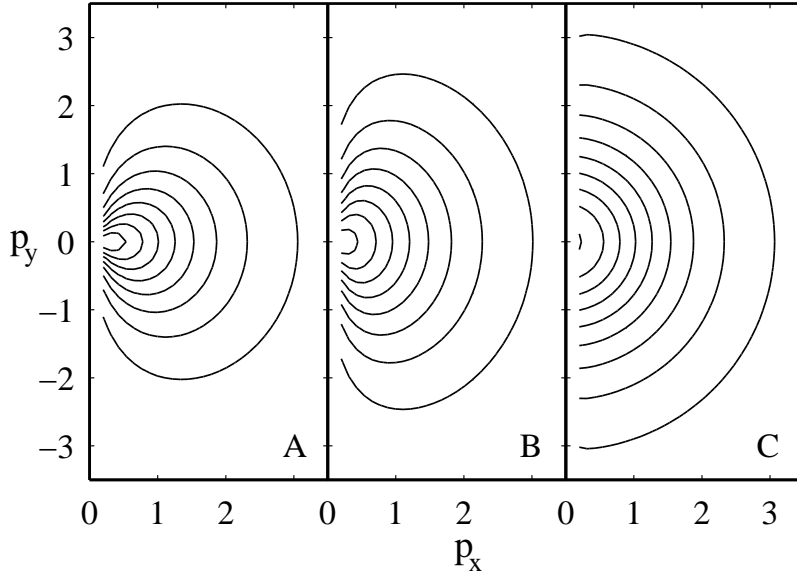}
\caption[]{\footnotesize
The freeze-out distribution, $f_{free}(x,\vec{p})$, in the rest frame
of the freeze out front (RFF) calculated from the model presented in
sect. \ref{six}. The momentum is plotted in units of [T], and $T=m$ is
assumed. Contour lines are given at values of 0.9, 0.8, 0.7, ... 
times the maximum of $f_{free}$.
Here the center of the Rest Frame of
the Gas (RFG) is at rest in RFF, $u_{RFG}^\mu=(1,0,0,0)$,
however, the Eckart and Landau flow velocities of the frozen
out matter do not vanish! The
figures A, B, C correspond to $x= 0.02 \lambda,\ 3\lambda,\ \infty$
respectively. At large distances from the initial point of the
freeze out process, $x \longrightarrow \infty$ (C),
the distribution becomes a cut J\"uttner distribution.
The earlier stages of the freeze out are, however, characterized by
asymmetric distributions, elongated in the freeze out direction, $x$.
This may lead to a large-$p_t$ enhancement, compared to the usual
J\"uttner assumption as freeze out distribution used in most previous
calculations.
}

%\label{fig:1}
\end{figure}

Let us take first the most simple kinetic model describing the evolution
of such a system. Starting from a fully equilibrated J\"uttner distribution
the two components of the momentum distribution develop according to the
coupled differential equations:
\beqar
\partial_x f_{int}(x,\vec{p})   dx &=& - \Theta(p^\mu d\sigma_\mu) 
                                   \frac{\cos \theta_{\vec{p}} }{\lambda}
           f_{int}(x,\vec{p})   dx,
\nonumber \\
\partial_x f_{free}(x,\vec{p})  dx &=& + \Theta(p^\mu d\sigma_\mu) 
                                   \frac{\cos \theta_{\vec{p}} }{\lambda}
           f_{int}(x,\vec{p})   dx.
\eeqar{kin-1}

Here the interacting component, $f_{int}$, will deviate from the
J\"uttner shape and the solution will take the form
\beq
f_{int}(x,\vec{p}) =  f_{Juttner}(x=0,\vec{p}) 
\exp \left[ - \Theta(p^\mu d\sigma_\mu) 
              \frac{\cos \theta_{\vec{p}} }{\lambda}  x \right].
\eeq{sol-11}
This solution is depleted in the forward $\vec{p}$-direction, particularly
along the $x$-axis.
Inserting it into the second differential equation above, leads
to the freeze out solution:
\beq
f_{free}(x,\vec{p}) =  f_{Juttner}(x=0,\vec{p}) 
\left\{1-\exp\left[-\Theta(p^\mu d\sigma_\mu)
\frac{\cos\theta_{\vec{p}}}{\lambda}x\right]\right\}.
\eeq{sol-12}
At 
$x \longrightarrow \infty$ 
this distribution will tend to the
cut J\"uttner distribution introduced in the previous section.
(see Figs. 1, 2, and 3.)
The remainder of the original J\"uttner distribution 
survives as $f_{int}$, even if
$x \longrightarrow \infty$.
In this model the particle density does not change with $x$, barely 
particles moving faster than the freeze out front 
(i.e. $p^\mu d\sigma_\mu  > 0$)
are transferred gradually from component $f_{int}$ to component $f_{free}$.
This is a highly unrealistic model, indicating that rescattering and 
re-thermalization should be taken into account in $f_{int}$. This would allow
particle transfer from the "negative momentum part" 
(i.e. $p^\mu d\sigma_\mu  < 0$)
of $f_{int}$ to $f_{free}$, which is not possible otherwise.

%   fig. 2

\begin{figure}[htb]

\insertplot{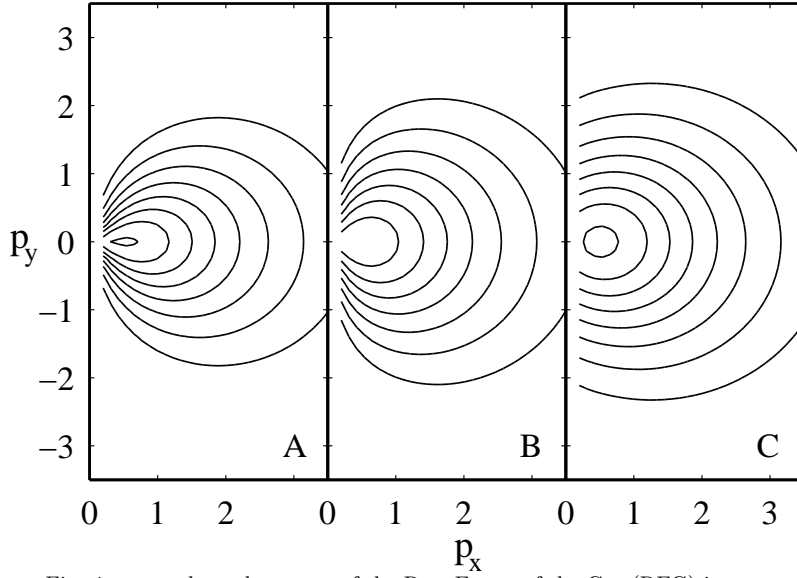}
\caption[]{\footnotesize
The same as Fig. 1, except here the center of the Rest Frame of
the Gas (RFG) is not at rest in RFF, $u_{RFG}^\mu=(\gamma ,0.5,0,0)$.
At large distances from the initial point of the
freeze out process, $x \longrightarrow \infty$ (C),
the distribution becomes a cut J\"uttner distribution, but less than
half of the distribution is cut off! Note that the boosted
J\"uttner distribution became Lorentz elongated and asymmetric
(see Fig. 2.10 of ref. \cite{Cs94}.)
The earlier stages of the freeze out, here also, are characterized by
asymmetric distributions, elongated in the freeze out direction, $x$.
}

\label{fig:2}
\end{figure}

\section{Freeze out distribution with rescattering}
\label{seven}

The assumption that the interacting part of the distribution remains
the distorted (after some drain) J\"uttner distribution, is of course
highly unrealistic. Rescattering within this component will lead to
re-thermalization and re-equilibration of this component. Thus the 
re-equilibration and the drain terms are in competition and they mutually
determine the evolution of the component, $f_{int}$.

To include the collision terms explicitly into the transport equations,
(\ref{kin-1}) leads to a combined set of integro-differential
equations. We can, however, take advantage of the  relaxation 
time approximation to simplify the description of the dynamics.

Then 
the two components of the momentum distribution develop according to the
coupled differential equations:
\beqar
\partial_x f_{int}(x,\vec{p})   dx &=& - \Theta(p^\mu d\sigma_\mu) 
                                   \frac{\cos \theta_{\vec{p}} }{\lambda}
           f_{int}(x,\vec{p})   dx
\nonumber \\
                                   &+& 
           \left[ f_{eq}(x,\vec{p}) -  f_{int}(x,\vec{p})\right]
           \frac{1}{\lambda'}   dx,
\nonumber \\
\partial_x f_{free}(x,\vec{p})  dx &=& + \Theta(p^\mu d\sigma_\mu) 
                                   \frac{\cos \theta_{\vec{p}} }{\lambda}
           f_{int}(x,\vec{p})   dx.
\eeqar{kin-2}
Here, the interacting component of the momentum distribution shows the 
tendency to approach an equilibrated, J\"uttner type, distribution with a 
relaxation length coefficient, $\lambda' \approx \lambda$.  Of course
due to the energy, momentum and conserved particle drain, this distribution,
$f_{eq}(x,\vec{p})$ is not the same as the initial J\"uttner distribution,
but its parameters, $n_{eq}(x)$, $T_{eq}(x)$ and $u^\mu_{eq}(x)$,
change as required by the conservation laws.

%   fig. 3

\begin{figure}[htb]

\insertplot{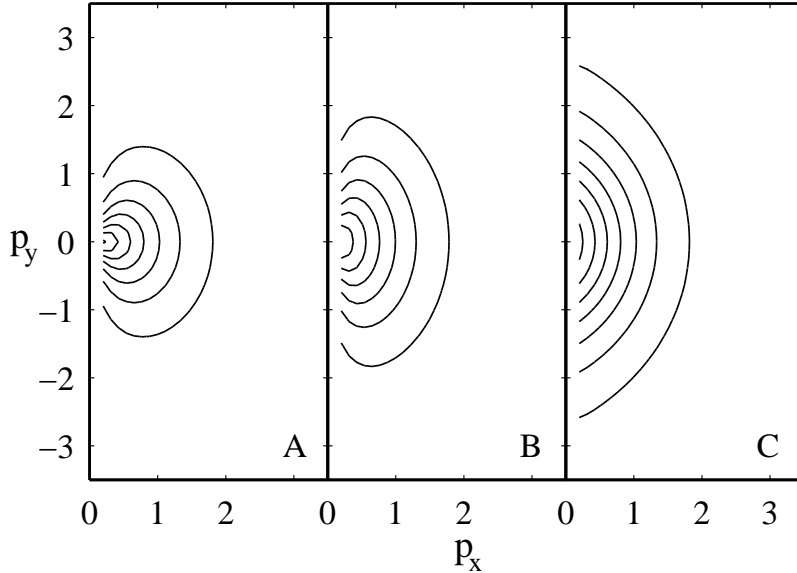}
\caption[]{\footnotesize
The same as Fig. 1, except here the center of the Rest Frame of
the Gas (RFG) is not at rest in RFF, $u_{RFG}^\mu=(\gamma ,-0.5,0,0)$.
At large distances from the initial point of the
freeze out process, $x \longrightarrow \infty$ (C),
the distribution becomes a cut J\"uttner distribution, but more than
half of the distribution is cut off!
The earlier stages of the freeze out, here also are characterized by
asymmetric distributions, but these are not 
elongated in the freeze out direction, $x$.
}

\label{fig:3}
\end{figure}

{\em Conservation Laws.} \ \ 
In this case the change of the conserved quantities 
caused by the particle transfer from component $int$ to component $free$
can be obtained in terms of the distribution functions as:
$$
d N_i^\mu = 
      -\frac{dx}{\lambda}\int\frac{d^3p}{p_0} p^\mu    
 \Theta(p^\mu d\sigma_\mu) \cos \theta_{\vec{p}}
   f_{int}(x,\vec{p})
$$
and
$$
d T_i^{\mu\nu} = 
      -\frac{dx}{\lambda}\int\frac{d^3p}{p_0} p^\mu p^\nu 
 \Theta(p^\mu d\sigma_\mu) \cos \theta_{\vec{p}}
f_{int}(x,\vec{p}).
$$
If we do not have collision or relaxation terms in our transport equation
then the conservation laws are trivially satisfied. If, however, collision or
relaxation terms are present these contribute, to the change of $T^{\mu\nu}$
and $N^\mu$, and this should be considered in the modified 
distribution function $f_{int}(x,\vec{p})$.

\paragraph{Immediate re-thermalization limit.}
As a first approximation to the solution of eqs. (\ref{kin-2}) let us assume 
that $\lambda' \longrightarrow 0$, i.e. we have immediate re-thermalization 
after every step $dx$. Thus the drain is always happening from a component of
shape $f_{eq}(x,\vec{p})$, with parameters,
$\hat{n}(x),\ T(x)$ and $u_{RFG}^\mu(x)$, and we can assume that 
$f_{int} = f_{eq}$ 
is of spherical J\"uttner form at any $x$ including both positive and
negative momentum parts. Above and henceforth the notation is similar to 
the one in \cite{AC98}: 
 $\tilde{n}=8\pi T^3e^{\mu /T}(2\pi \hbar )^{-3}$, \ 
$a={\frac mT}$, \ so that $\hat{n}(\mu ,T)=\tilde{n}a^2K_2(a)/2$ is the 
invariant scalar density of the symmetric massless J\"{u}ttner gas, 
$b=a/\sqrt{1-v^2}$%
, \ $v=d\sigma _0/d\sigma _x$, \ $A=(2+2b+b^2)e^{-b}$, \ and 
\[
{\cal K}_n(z,w)\equiv \frac{2^n (n)!}{(2n)!}\
z^{-n}\!\!\int_w^\infty \!\!\!dx\ (x^2-z^2)^{n-1/2}\ \ e^{-x}\ , 
\]
i.e. ${\cal K}_n(z,z)=K_n(z)$. 

In this case the change of conserved quantities due to particle drain or 
transfer can be  evaluated for an
infinitesimal $dx$. 
We assume that the 3-flow is normal to the
freeze out surface, and for simplicity we assume $v > 0$.
In this case the change of the conserved particle currents
in the RFF is given by 
$$
\begin{array}{rll} 
dN_i^0
=&-\frac{dx}{\lambda}\frac{\tilde{n}}{4v^2\g^2}
\left[
bK_1(b)+b(3v^2-1)\g^2(2K_1(a)-{\cal K}_1(a,b))
\right.+\nonumber\\ & & \\ 
&+\left.
\g v^2b^2(2K_0(a)-{\cal K}_0(a,b))+2v^3\g^3(b+1)e^{-b}
\right]\nonumber\ ,\\ & & \\
dN_i^x
=&-\frac{dx}{\lambda}\frac{\tilde{n}}{4v^3\g^3}
\left\lbrace
v^2(3v^2-1)\g^3b(2K_1(a)-{\cal K}_1(a,b))+
\right.\nonumber\\ & & \\  
&+(2+v^4\g^2b^2)(2K_0(a)-{\cal K}_0(a,b))-
\nonumber\\ & & \\ 
&\left.-2K_0(b)+2v\g^2e^{-b}[v^2\g^2(b+1)+v^2b-1]
\right\rbrace ,
\label{dni}
\end{array}
$$
and for the change of the energy - momentum tensor in the RFF we obtain that
$$
\begin{array}{rll}
dT_i^{00}=&-\frac{dx}{\lambda}\frac{\tilde{n}T}{4v^2\g^2}
\left\lbrace
  v^2\g^2b^2(3+v^2)(2K_2(a)-{\cal K}_2(a,b))+
\right.
\\ & & \\
 &+(v^2b^2-v^2-1)\g b(2K_1(a)-{\cal K}_1(a,b))-
   \\ & & \\
&-b^2(2K_0(a)-{\cal K}_0(a,b))+
\frac{a}{\g}K_1(b)+a^2K_0(b)+
 \\ & & \\
&\left.+v\g^2e^{-b}\!\left[ (1{+}3v^2)\g^2\!A(b)-(2{+}v^2b^2)(b{+}1)+
v^4(1{+}\frac{v^2}{3})\g^2b^3\right]\right\rbrace ,
\\ & & \\
dT_i^{0x}=&-\frac{dx}{\lambda}\frac{\tilde{n}T}{4}
\left\lbrace 
\frac{1+3v^2}{v}b^2(2K_2(a)-{\cal K}_2(a,b))+
\right.
\\ & & \\
 &+vab^2(2K_1(a)-{\cal K}_1(a,b))+
\left[ 
v^2\g^2b
\left(
- a^2+\frac{1+3v^2}{3v}b^2
\right)
-b^2+
\right.
\\ & & \\
& \left.
\left.
+(v^2+3)\g^2A(b)
\right]
e^{-b}
\right\rbrace
-\frac{2T}{v\g}dN_i^0
\ ,
\end{array}
$$
%\pagebreak
$$
\begin{array}{rll}
dT_i^{xx}=& -\frac{dx}{\lambda}\frac{\g^2\tilde{n}T}{4v}
\left\lbrace
v(3+v^2)a^2(2K_2(a)-{\cal K}_2(a,b))+\right.
\\ & & \\
&+v^3a^3(2K_1(a)-{\cal K}_1(a,b))+
\left[
\frac{v^4}{3}(3+v^2)b^3+
\right.
\\ & & \\
&\left. \left. 	
+a^2(v^4b-1)+(3v^2+1)A(b)
\right]	
e^{-b}
\right\rbrace
-\frac{3T}{v\g}dN_i^x\ ,
\\ & & \\
dT_i^{yy}=& {-}\frac{dx}{\lambda}\frac{\tilde{n}T}{8v^2\g^2}
\left[
-(v^2+1)a(2K_1(a)-{\cal K}_1(a,b))-
\right.
\\ & & \\
&\left.
-v^2a^2\left(2K_0(a){-}{\cal K}_0(a,b)\right)
+\frac{a}{\g}K_1(b)-2v(b{+}1)e^{-b}
\right]+\frac{3T}{2v\g}dN_i^x 
\end{array}
$$
and $dT_i^{zz} = dT_i^{yy}$.
Note that in RFF the flow velocity of the re-thermalized component is
$u_{i,RFG}^\mu(x) = \gamma_\sigma(x)\ (1,v(x),0,0)|_{RFF}$, where
$\gamma_\sigma = 1/\sqrt{1-v^2}$.

The new parameters of distribution $f_{int}$, 
after moving to the right by $dx$ can be obtained from
$dN_i^\mu$ and $dT_i^{\mu\nu}$.
The conserved particle density of the re-thermalized spherical
J\"uttner distribution after a step $dx$ is 
$$
\hat{n}_i(x+dx) = \hat{n}_i(x) + d\hat{n}_i(x)
 = 
\sqrt{ N_i^\mu(x+dx) N_{i,\mu}(x+dx) }
$$
where the expressions are invariant scalars.
%, thus
%these can be calculated in any reference frame. In the RFG frame
%$$
%N_i^\mu(x) = ( \hat{n}_i(x), 0, 0, 0 )|_{RFG}  
%           =   \hat{n}_i(x) \ u_{i,RFG}^\mu(x)
%$$ 
%and  the quantities
%$
%dN_i^\mu(x) 
%$ 
%are calculated above. Since 
%$
%N_i^\mu(x+dx)$ $  = $ $ N_i^\mu(x) + dN_i^\mu(x)
%$
%in any reference frame, we can calculate
%$$
%\hat{n}_i(x+dx) =
%\sqrt{ ( N_i^0(x) + dN_i^0(x) )^2  - ( N_i^x(x) + dN_i^x(x) )^2 }\,, 
%$$
%where $N_i^\mu$ and $dN_i^\mu$
%are in the RFG  measured at $x$. 
%Thus the proper density  at $x+dx$ is
%%$$
%\hat{n}_i(x+dx) =  
%\hat{n}_i(x) 
%\sqrt{ \left( u_{i,RFG}^0(x) + \frac{dN_i^0(x)}{\hat{n}_i(x)} \right)^2  
%     - \left( u_{i,RFG}^x(x) + \frac{dN_i^x(x)}{\hat{n}_i(x)} \right)^2 } 
%$$
%$$
%\approx 
%\hat{n}_i(x) 
% +  u_{i,RFG}^0 dN_i^0(x)  
% -  u_{i,RFG}^x dN_i^x(x)  
% + \Theta(dx^2) \,,
%$$
%where we  took advantage of the normalization of the flow velocity and
%neglected terms second order in $dx$. Thus 
After straightforward calculation
the differential equation
describing the change of the proper particle density is:
\beq
d\hat{n}_i(x) =  u_{i,RFG}^\mu(x)\ dN_{i,\mu}(x) \,. 
\eeq{dnx}
Although this covariant
equation is valid in any frame, we can
calculate it in the RFF, where 
the values of $dN_i^\mu$s were given above.
Note again, that the particle drain from $f_{int}(x)$, 
described by $dN_i^\mu$ is  constrained to the "positive part"
in the momentum space, but after re-thermalization we attribute this to
the 
change  in the complete spherical J\"uttner distribution, $f_{int}(x+dx)$. 
Thus, in order to conserve momentum, we have to obtain a decreased
Eckart flow velocity after the infinitesimal particle drain.

For the re-thermalized interacting component
Eckart's flow velocity is the velocity of the RFG, which changes with $x$,
so we can actually denote this frame as RFG$(x)$.
For the spherical J\"uttner distribution  
the Landau and Eckart flow velocities are the same,
$
u^\mu_{i,E,RFG}(x) = u^\mu_{i,L,RFG}(x) 
= u^\mu_{i,RFG}(x) 
$.
Thus we can evaluate the flow velocity $u_{i,RFG}^\mu(x+dx)$:
$$
u^\mu_{i,RFG}(x+dx)  =    N_i^\mu (x+dx) / \sqrt{ N_i^\mu N_{i,\mu} }\,,
$$
which leads to the following covariant  expression
%$$
%u^\mu_{i,E,RFG}(x+dx) = \frac{
% \hat{n}_i(x) u_{i,RFG}^\mu(x) + dN_i^\mu(x)   
%}{ 
% \hat{n}_i(x) + u_{i,RFG}^\nu(x) dN^\mu_i(x)   
%}
%$$
%$$
%\approx 
% u_{i,E,RFG}^\mu(x) 
%\left( 
%1 - \frac{u_{i,E,RFG}^\nu(x) dN_{i,\nu}(x)}{\hat{n}_i(x)} 
%\right) +
%\frac{ dN_{i,\mu}(x)}{\hat{n}_i(x)} \,, 
%$$
%Thus  the change of the Eckart flow velocity 
%of the interacting particle component is:
\beq
du^\mu_{i,E,RFG}(x)=\Delta^{\mu\nu}_i(x)\  \frac{dN_{i,\nu}(x)}{\hat{n}_i(x)}\,,
\eeq{duex}
where
$
\Delta^{\mu\nu}_i(x) = g^{\mu\nu} -  u_{i,RFG}^\mu(x)\, u_{i,RFG}^\nu(x)\,, 
$\ \ 
is a projector to the plane orthogonal to
$
u^\mu_{i,RFG}(x)  
$.
This equation is valid in any reference frame, nevertheless we know the 
four-vectors on the r.h.s. in the RFF explicitly.
Then the new flow velocity of the matter 
evaluated according to Eckart's definition is
$
u^\mu_{i,E,RFG}(x+dx) $ $ = $ $  
u^\mu_{i,RFG}(x)$ $ + $ $  du_{i,E,RFG}^\mu(x) 
$.

%   fig. 4

\begin{figure}[htb]
\vskip -0.95cm
\insertplot{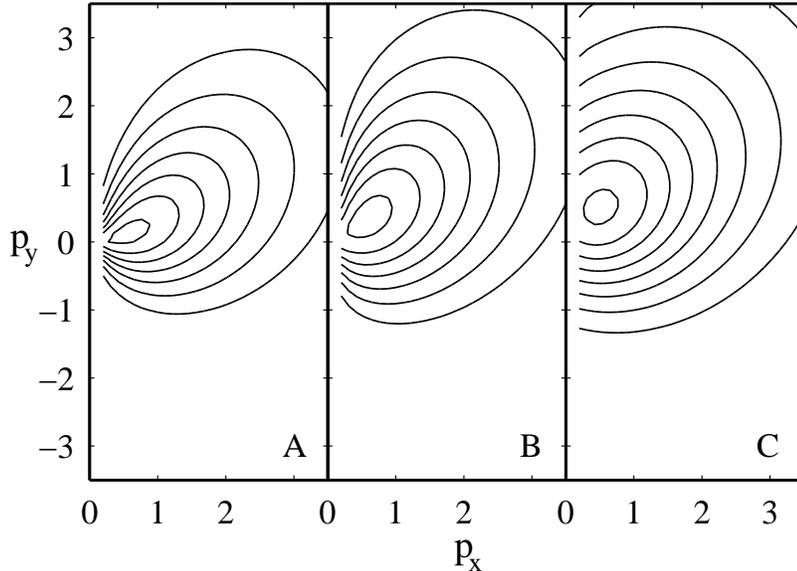}
\caption[]{\footnotesize
The same as Fig. 1, except here the center of the Rest Frame of
the Gas (RFG) is not at rest in RFF, $u_{RFG}^\mu=(\gamma ,0.5,0.5,0)$.
At large distances from the initial point of the
freeze out process, $x \longrightarrow \infty$ (C),
the distribution becomes a cut J\"uttner distribution, which is
not centralized in $p_y$ and less than
half of the distribution is cut off!
The earlier stages of the freeze out, here also are characterized by
distributions asymmetric both in the directions $p_x$ and $p_y$, 
and these are also elongated in the direction of the freeze out flow
velocity, $u_{RFG}^\mu$.
}

\label{fig:4}
\end{figure}

To get the temperature and the change of Landau's flow velocity, we have to
analyze the change of the energy momentum tensor.
Before the particle drain the energy - momentum tensor at $x$ in the RFG is 
diagonal, 
$T^{\mu\nu}_i(x) = {\rm diag}(e_i,P_i,P_i,P_i)|_{RFG(x)}$
while in the RFF 
$T^{\mu\nu}_i(x) $ $=$ $ 
\left[ (e_i + P_i) \right.$ $ u_{i,RFG}^\mu u_{i,RFG}^\nu(x) 
$ 
$ 
\left. - P_i g^{\mu,\nu}\right] |_{RFF(x)}
$.
Adding the
drain terms, $dT^{\mu\nu}_i(x)$, to this arising from the freeze out while
we move to the right  by $dx$, yields 
$T^{\mu\nu}_i(x+dx)$ 
which will not be diagonal in the RFG$(x)$ and the 
pressure part will not be isotropic. We can Lorentz transform
this to another frame which diagonalizes 
$T^{\mu\nu}_i(x+dx)$.
This means to find the Landau flow velocity of the new system,
$u^\mu_{i,L,RFG}(x+dx)$ in the original RFG$(x)$. After
a straightforward diagonalization, a somewhat tricky algebra and 
neglecting second and higher order terms
we arrive at the covariant expression%
\footnote{
Let the energy-momentum tensor of a system be $T^{\mu\nu}$. The energy and
momentum flow is characterized by the Landau flow velocity, a unit four
vector, 
$u_\mu$. We are looking for a relationship between the infinitesimal
change of
the flow velocity $du_\mu$ and the corresponding shift in the
energy-momentum 
tensor $dT^{\mu\nu}$.
We introduce the projector, $\Delta^{\mu\nu}=g^{\mu\nu}-u^{\mu}u^{\nu}$,
with the properties\,\cite{Cs94}
$
\Delta^{\mu\nu}u_{\nu}=0
$
and
$
du_\mu=\Delta_\mu^\nu du_\nu 
$
since
$
u_\mu du^\mu=0\,.
$
The Landau flow velocity is parallel to the flow of the momentum. Thus
$
u_\mu=Const.\times T_\mu^\nu u_\nu\,,
$
therefore 
$
\Delta_{\rho\mu}T^{\mu\nu} u_\nu=0\,.
$
We differentiate the above equation and take into consideration the
identities
$
e\equiv u_\mu T^{\mu\nu} u_\nu
$
and
$
\Delta^\mu_\rho T^{\rho\sigma} \Delta^\nu_\sigma=-P\Delta^{\mu\nu}\,,
$
where $e$ and $P$ are the energy density and pressure of the
dissipationless,
fully equilibrated fluid.
Then using the properties of $\Delta^{\mu\nu}$ we get
$
du_\rho(e+P)+u_\rho du_\mu
T^{\mu\nu}u_\nu=\Delta_{\rho\mu}dT^{\mu\nu}u_\nu\,.
$
Since the flow velocity and the momentum flow are parallel the second
term on
the l.h.s vanishes. Thus the equation describing the change of Landau's
flow
velocity becomes
$
du_\rho=\Delta_{\rho\mu}dT^{\mu\nu}u_\nu/(e+P)\,.
$
}
%%%%%%%%%%%%%%%%%%%
%% End of footnote%
%%%%%%%%%%%%%%%%%%%
\beq
du^\mu_{i,L,RFG}(x) = \frac{ 
\Delta^{\mu\nu}_i(x)\ \ dT_{i,\nu\sigma}\ \ u^\sigma_{i,RFG}(x)}{e_i + P_i}.
\eeq{dulx}
Although, 
for the spherical J\"uttner distribution  
the Landau and Eckart flow velocities are the same,
the change of this flow velocity when calculated from the baryon current
and from the energy current are different
$$
du^\mu_{i,E,RFG}(x) \ne  du^\mu_{i,L,RFG}(x) \,.
$$
This is a clear consequence of the asymmetry caused by the freeze out
process as we pointed out already at the discussion of the properties
of the cut J\"uttner distribution. Unfortunately, this also illustrates the 
weakness of our assumption on the complete re-thermalization to a
spherical J\"uttner distribution, because we cannot choose the
correct velocity change: If we choose $du^\mu_E$ as the new velocity
of the (spherical J\"uttner distribution) $f_{int}(x+dx)$, 
then we violate the momentum conservation in our model,
on the other hand if we choose $du^\mu_L$, then we violate the baryon
current conservation! Thus a spherical (or even elliptic) distribution
cannot be fitted to the freeze out drain, and we would have to use an
ansatz, which has (in addition) an asymmetry in the $x$ direction
(i.e., an egg shape), for the distribution $f_{int}$.

Being aware of this weakness of the model, we nevertheless, maintain
the assumption of spherical J\"uttner shape for $f_{int}$ for the sake of
simplicity. We can choose the flow velocity change then according
to the physical problem. For example for  the freeze out of baryon free plasma
this problem does not occur, and we have to choose $du^\mu_L$.

The last item is to determine the change of the temperature parameter
of $f_{int}$.  From the relation
$
e\equiv u_\mu T^{\mu\nu} u_\nu
$
we readily obtain the expression for the change of energy density
\beq 
de_i(x) =u_{\mu,i,RFG}(x) \ \ dT^{\mu\nu}_i(x) \ \  u_{\nu,i,RFG}(x) \,,
\eeq{dedx}
and from the relation between the energy density and the
temperature (see Chapter 3 in ref.\cite{Cs94}), we can obtain
the new temperature at $x+dx$.
Fixing these parameters we fully determined the spherical J\"uttner
approximation for $f_{int}$. With this ansatz the pressure asymmetry
and pressure balance cannot be realized, thus our model will
be only a rather approximate description of the freeze out process.

Nevertheless, we can draw some preliminary conclusions about the
development of the kinetic distribution during freeze-out.

\section{Conclusions}

We turned to the problem of estimating the freeze out distribution.
Obviously the real freeze out distribution depends strongly on the details
of the freeze out (and hadronization) dynamics. In heavy ion reactions, the
curvature of the freeze out surface and the conditions varying in time
do affect the freeze out distribution, nevertheless, as a first step, we
assumed that the process is stationary and the curvature of the front 
is negligible.  These approximations are extreme, but still enable us
to draw some preliminary conclusions.

Following the lines and ideas presented in ref. \cite{AC98},
the first simple kinetic freeze out model reproduces the cut J\"uttner
distribution as the limiting distribution, $f_{free}$, after complete freeze
out at large distances. However, the model at the same time leads to
unrealistic consequences, namely that the interacting part of the
distribution, $f_{int}$ also survives fully, as the other part of the
J\"uttner distribution.  Thus having both components at the end in this
model, the physical freeze out is actually not realized.  This turns out to
be a consequence of the fact that the effect of rescattering and
thermalization in the interacting part of the distribution was ignored. 

In an improved but still rather approximate kinetic
freeze out model which takes rescatterings into account,
the interacting component is assumed to be instantly re-thermalized
taking a spherical J\"uttner shape at each time step with changing parameters.
The model leads to a set of coupled differential equations 
(\ref{dnx},\ref{duex},\ref{dulx},\ref{dedx}). 
Equations (\ref{duex}) and (\ref{dulx}) 
can be used in some combined form, or one of them can be selected which fits
the physical situation the best. Then the three parameters of the
interacting component, $f_{int}$, can be obtained in each time step 
analytically (considering ${\cal K}_n(x,y)$ an analytic function). 

Now the density of the interacting component will gradually decrease 
and disappear according to eq. 
(\ref{dnx}), the flow velocity will also decrease in both cases, (\ref{duex})
or (\ref{dulx}), because only forward going particles freeze out, and
the energy density will decrease also according to eq. (\ref{dedx}).
Thus, the initial contribution to $f_{free}$ at small $x$ will
resemble the distribution shown in Fig. 2A, then as $x$ increases
and the velocity decreases it will become to similar to Fig. 1B,
while at the final stages it will approach Fig. 3C. As a consequence
the integrated distribution will not resemble a cut J\"uttner
distribution.

Thus the arising post freeze out distribution, $f_{free}$ will be a 
superposition of cut J\"uttner type of components, from a series of gradually
slowing down J\"uttner distributions. This will lead to a comet
shaped final momentum distribution, with a more dominant leading head
and a tail.  In these rough models a large fraction ($\sim 95\%$)
of the matter is frozen out by $x=3 \lambda$, thus the distribution
$f_{free}$ at this distance can be considered as a first estimate
of the post freeze out distribution.  One should also keep in mind that 
the models presented here  do not have realistic behavior in the limit 
$x \longrightarrow \infty$, due to their one dimensional character.
Nevertheless, this improved model with rescattering enables
complete freeze out (unlike the simpler model in sect. \ref{six}
where only the originally forward moving particles freeze out
even at large distances).

In case of rapid hadronization of QGP and 
simultaneous freeze out, the idealization of a freeze out hypersurface
may be justified, however, an accurate determination of the post freeze
out hadron momentum distribution would require a nontrivial dynamical
calculation.

\section*{Acknowledgement}

Enlightening discussions  with
F. Grassi, Y. Hama and T. Kodama
are gratefully acknowledged.
This work is supported in part by the Research Council of Norway (programs for
nuclear and particle physics, supercomputing, and free projects).
Cs. Anderlik, L.P. Csernai and Zs.I. L\'az\'ar are thankful for the
hospitality extended to them by the Institute for Theoretical Physics
of the University of Frankfurt where part of this work was done. 
Cs. Anderlik and Zs.I. L\'az\'ar are also indebted to the 
Theoretical Section of GSI, Darmstadt for their help.

\end{document}